\documentclass[12pt]{iopart}
\usepackage{graphicx}
\usepackage{setspace}
\usepackage{bm}
\usepackage{enumerate}
\usepackage{amssymb}
\usepackage{color}

\begin{document}

\title{Multiband dynamics of extended harmonic generation in solids under ultraviolet injection}

\author{Yue Lang$^1$, Zhaoyang Peng$^1$, Zengxiu Zhao$^1$\footnote{Corresponding author: zhaozengxiu@nudt.edu.cn}}

\address{$^1$ Department of Physics, National University of Defense Technology, Changsha, P. R. China}
\vspace{10pt}

\begin{abstract}

Using one-dimensional semiconductor Bloch equations, we investigate the multiband dynamics of electrons in a cutoff extension scheme employing an infrared pulse with additional UV injection. An extended three-step model is firstly validated to play a dominant role in emitting harmonics in the second plateau. Surprisingly, further analysis employing the acceleration theorem shows that though harmonics in both the  primary and secondary present positive and negative chirps, the positive (negative) chirp in the first region is related to the so-called short (long) trajectory, while that in the second region is emitted through `general' trajectory, where electrons tunnelling earlier and recombining earlier contribute significantly. The novel characteristics deepen the understanding of high harmonic generation in solids and may have great significance in attosecond science and reconstruction of band dispersion beyond the band edge.

\end{abstract}

\maketitle

\section{Introduction}

Highly nonlinear interaction of multicycle laser pulse with gaseous targets enables the generation of high harmonics (HHG) and lays the foundation of attosecond science\cite{Krausz2009attosecond}. The process of up-converting the fundamental frequency can be well described with the semiclassical three-step model including tunneling, acceleration and recombination\cite{Corkum1993plasma,Lewenstein1994theory} and has been extensively studied during the past three decades\cite{Ferray1988multiple,McPherson1987studies,Paul2001observation,Hohenleutner2015real,Popmintchev2012bright,Zhang2013elliptical}. Due to the rich information carried by HHG from the microcosm, it has been widely applied to the detection of structures and control of ultrafast dynamics with unprecedented spacial and temporal resolutions\cite{Kienberger2002steering,Klunder2011probing,Drescher2002time,Maquet2014attosecond,Midorikawa2022,Uiberacker2007attosecond,Zhang2012synchronizing,Huang2015joint,Huang2021ultrafast,Zhao2021strong,Dong2020time}. The cutoff in gases is theoretically and experimentally given by $\hbar\omega_{\mathrm{cutoff}}\approx I_p+3.17U_p\propto I\lambda^2$, illustrating the possibility of extending the energy of harmonic photons well into extreme ultraviolet and even ``water window" region employing longer wavelengths\cite{Li201753,Li2019double,Ren2018attosecond,Wang2020generation}.

HHG from solids has also been scrutinized theoretically during the same period\cite{Plaja1992high,Faisal1997floquet,Liu2017spacial,Liu2019role}, partially due to its promise of intenser and higher flux of emission with much denser electrons. However, the experimental observation was reported only recently in Zinc Oxide\cite{Ghimire2011observation}, benefiting from longer wavelength of the incident pulse to circumvent radiation damage in condensed materials. Mechanism of solid HHG still remains under intense debate and is partially shed light on in pump-probe experiments\cite{Vampa2015linking,Vampa2014theoretical,Vampa2020attosecond}, where the rescattering model is found to predict satisfyingly the observations. The high efficiency of HHG in solids replenishes the development of table-top attosecond sources\cite{Ndabashimiye2016solid} and enables petahertz-bandwidth signal manipulation through ultrafast control of current injection\cite{Schultze2013controlling,Schiffrin2013optical}. Also, deeper insights into the mechanism of HHG in solids hold the promises of reconstruction of band structures\cite{Vampa2015all}, potential field imaging\cite{Lakhotia2020laser}, detection of topological transition\cite{Silva2019topological}, etc. Different from the counterpart in gases, the solid-state cutoff is observed to scale linearly with the electric field\cite{Ghimire2011observation} and has been demonstrated to depend on the band dispersion\cite{Nourbakhsh2021high}, while direct relevance with driving wavelength is still under debate\cite{Liu2017time}. As a result, potential applications call for deeper understanding of solid HHG and new methods to extend the cutoff.

Various schemes have been explored intensively to increase the cutoff and intensity of harmonics in solids, such as the employment of solid Ar (or Kr)\cite{Ndabashimiye2016solid}, the consideration of pre-acceleration\cite{Li2019reciprocal}, properly-doped semiconductor\cite{Nefedova2021enhanced,Yu2019enhanced}, etc. On the other hand, different compositions of incident pulses, such as two color or even three color, have been explored where the relative intensity, delay, carrier-envelope phase (CEP), and pulse shape can be adjusted to achieve better characteristics of resulting harmonic spectra\cite{Song2020enhanced,Navarrete2019crystal,Shao2020quantum,Uzan2020attosecond,Sun2021cutoff}. In solids, further extend of harmonics to higher energy region requires intenser incident pulse to drive direct transitions from the highest valence band to higher-lying conduction bands. Due to the large gaps, however, the incident pulse should be intense enough to achieve significant population, which can possibly cause irreversible damage to materials. An appearing bypass is coincidence of infrared (IR) and properly-delayed ultraviolet pulses, hereinafter referred to as IR+UV scheme.   We recently have proposed a model to extend the cut-off energy of harmonic generated from solids by injecting both the IR and UV pulses \cite{Yue2022}. In this study, we further verify the extended three-step model and discuss the effects of multiple bands. In particular, we  examine the difference in the interband dynamics of the two plateaus. The paper is organized as follows: The theoretical model and approaches are given in the methods section; Then, we present the extended three-step model and focus on the multiband dynamics in the extended plateau; Finally, we summarize the findings in the conclusion part.

\section{Methods}

The dynamics of electrons in laser solid interaction can be given by the semiconductor Bloch equations (SBE)\cite{Yue2020structure,Yue2022introduction,Foldi2017gauge}:
\begin{eqnarray}\label{SC-SBE}
i\frac{\partial}{\partial t}{\rho}_{mn}^{\mathbf{k}}\left( t \right) =&\left[ E_{m}^{\mathbf{k'}}-E_{n}^{\mathbf{k'}}-\frac{i\left( 1-\delta _{mn} \right)}{T_2} \right] \rho _{mn}^{\mathbf{k}}\left( t \right)\\
&+\mathbf{F}\left( t \right) \cdot \sum_l{\left[ \mathbf{d}_{ml}^{\mathbf{k'}}\rho _{ln}^{\mathbf{k}}\left( t \right) -\mathbf{d}_{ln}^{\mathbf{k'}}\rho _{ml}^{\mathbf{k}}\left( t \right) \right]},
\end{eqnarray}
where $\mathbf{k}$ is the crystal momentum, $m$ (or $n$) is the band indice,  $\rho_{mn}$ is the density matrix element, with $m=n$ representing interband polarization and $m\neq n$ population, $E$ is the band dispersion, $\mathbf{F}(t)$ is the electric field, $\mathbf{d}$ is the transitional dipole moment (TDM), $T_2$ is the empirical dephasing time and $\mathbf{k}'=\mathbf{k}+\mathbf{A}(t)$ is the crystal momentum in the moving frame. Note that $T_2$ is set to infinity unless otherwise stated and the influence of shorter dephasing time \cite{Peng2022decoherence} will be illustrated later.

The total current can be split into the interband and intraband contributions:
\begin{eqnarray}
\mathbf{J}_{\mathrm{inter}}\left( t \right) =-\int_{BZ}{d}\mathbf{k}\sum_{m\ne n}{\rho_{mn}^{\mathbf{k}}\mathbf{p}_{nm}^{\mathbf{k}}}, \label{Jinter}
\\
\mathbf{J}_{\mathrm{intra}}\left( t \right) =-\int_{BZ}{d}\mathbf{k}\sum_{m=n}{\rho_{mn}^{\mathbf{k}}\mathbf{p}_{nm}^{\mathbf{k}}}, \label{Jintra}
\end{eqnarray}
where the momentum matrix element (MTM) $\mathbf{p}_{nm}^{\mathbf{k}}$ is related to the TDM $\mathbf{d}_{nm}^{\mathbf{k}}$ through canonical relation\cite{Lindefelt2004choice}
\begin{equation}\label{DP}
\mathbf{d}^{\mathbf{k}}_{nm}= 
\frac{-i\mathbf{p}^{\mathbf{k}}_{nm}}{E^{n}_{\mathbf{k}}-E^{m}_{\mathbf{k}}}\ (\mathrm{for}\ m \ne n).
\end{equation}
The calculation of one dimensional band dispersion and TDM are described in other works\cite{Yue2020structure} for model crystal ZnO along $\Gamma$-M.

The harmonic spectrum is calculated by Fourier transformation
\begin{equation}\label{FFT}
\mathrm{Y}(\omega) = \left| \int_0^T\left[\frac{\partial \mathbf{J}(t)}{\partial t}w(t)\right]e^{i\omega t}dt \right|^2,
\end{equation}
where $T$ is the calculation time, $w(t)$ is the absorption function given by
\begin{eqnarray}
w\left( t \right) =\left\{ \begin{array}{l}
1\ \ \ \ \ \ \ \ \ \ \ \ \ ,\ t<T_0\\
e^{\frac{-4\ln2\left( t-T_0 \right) ^2}{\sigma^2}} ,\ t\ge T_0\\
\end{array} \right. ,\label{FFT_ABS}
\end{eqnarray}
with $T-T_0\approx10$fs. Fourier transformation in Eq. (\ref{FFT}) is conducted using Fast Fourier Transformation (FFT) algorithm.

Time-frequency analysis of the total current is calculated by continuous wavelet transformation
\begin{equation} \label{cwt}
\mathrm{S}(t, \omega)=\left|\frac{1}{\sqrt{a}} \int_{-\infty}^{\infty} \frac{\partial }{\partial t}\mathbf{J}\left(t^{\prime}\right) \alpha \left(\frac{t^{\prime}-t}{a}\right) d t^{\prime}\right|^{2},
\end{equation}
where $a$ is the scale factor, $\alpha(x)$ is the mother wavelet, i.e. Morlet wavelet in our case with the time-bandwidth set to 60 to achieve better resolution in time and space.

\section{Results and Discussion}

\begin{figure}[htb]
\includegraphics[width=10cm]{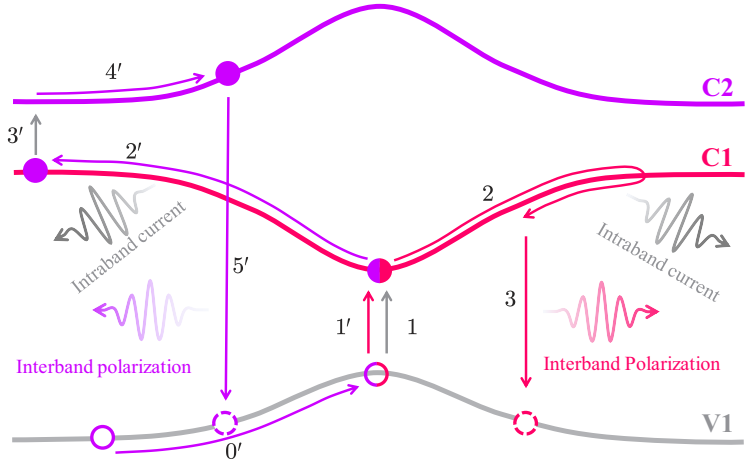}
\centering
\caption{ \label{Model}
Sketch of the extended three-step model and the general recollision model. Left panel: acceleration ($0'$), injection ($1'$), acceleration ($2'$), tunneling ($3'$), acceleration ($4'$) and recombination ($5'$); Right panel: tunneling ($1$), acceleration ($2$) and recombination ($3$). In the extended three-step model, excited electrons on band C2 originate far from the BZ center (magenta solid circle) on band V1, followed by injection to band C1 through properly-delayed UV pulse.  
}
\end{figure}

The general recollision model in two-band approximation \cite{Vampa2014theoretical,Vampa2015linking,Ghimire2019high} is shown on the right panel of Fig. \ref{Model}. Near the extreme of electric field, electrons locating initially around high symmetry points are partially excited to the lowest conduction band and form electron hole pairs. Then, electrons (holes) are accelerated by the laser field, with harmonics emitted from intraband current. Finally, when the electric field reverses its direction, excited electrons are driven back towards the holes and complete recombination with emission of harmonics, commonly referred to as the interband mechanism. For direct band-gap materials like ZnO\cite{Ghimire2011observation} along specific crystal direction, excited electrons locate generally near the center of Brillouin zone (BZ), i.e. $\mathbf{k}=0$, which impedes the following propagation to the band edge to be further pumped to higher-lying band C2.

In order to extend the cutoff, an additional ultraviolet (UV) pulse can be introduced and the dynamics of electrons are shown on the left panel of Fig. \ref{Model}. Electrons locating initially far from the BZ center are driven by the IR field (step $0'$) to the center, followed by excitation to the lowest conduction band C1 through one-photon absorption at the instant of UV injection (step $1'$). Then these electrons can be accelerated closer to the BZ boundary (step $2'$) and tunnel to band C2 (step $3'$) for further acceleration (step $4'$). Finally, the electrons on C2 recombine with the hole pairs and emit high energy photons (step $5'$). Note that though the injection scheme has been employed by Sun {\it et al}\cite{Sun2021cutoff}, the motion of electrons in their work is limited to only two bands, which differs from the extended three-step model.

\begin{figure}[htb]
	\includegraphics[scale=0.75]{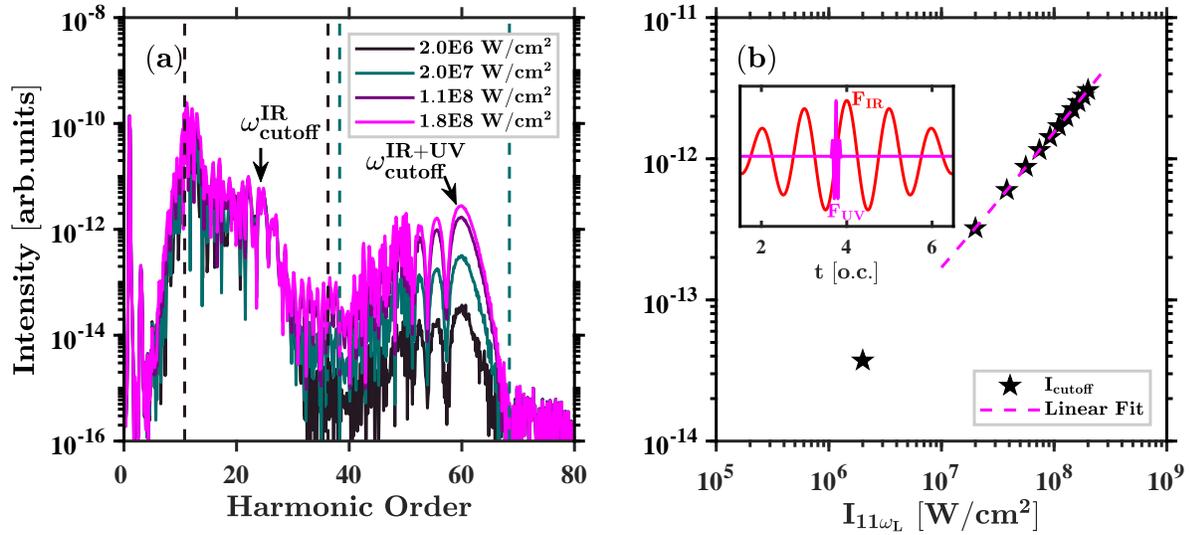}
	\centering
	\caption{ \label{HHG}
		(a) Harmonic spectrum under various injection intensities with IR fixed at $2\times10^{11}\mathrm{W/cm}^2$. The dashed lines represent energy gaps between the highest valence band and the lowest ($11\omega_L$ to $36\omega_L$) and second lowest ($38\omega_L$ to $68\omega_L$) conduction band, respectively; (b) The dependence of cutoff intensity on the injection intensity (dark pentagrams). Inset shows the separately-rescaled pulse shape of IR (red) and UV (magenta) fields. Absorption of one UV photon is indicated by the linear fit (magenta dashed line).
	}
\end{figure}

The pulse shape of both IR and UV is given by
\begin{eqnarray}
F(t) = F_{0} \cos^2\left[\pi\left(\frac{t-t_{\mathrm{delay}}}{T}-\frac{1}{2}\right)\right]\sin\left[\omega\left(\frac{t-t_{\mathrm{delay}}}{T}-\frac{1}{2}\right)\right],
\end{eqnarray}
with $F_0$ the electric amplitude, $T$ the total duration, $\omega$ the angular frequency and $t_{\mathrm{delay}}$ the delay time. The IR pulse is centered at 3200 nm ($\omega_L=0.39$ eV), with a total duration of 8 cycles (8$T_L$) and an intensity of $2\ \times10^{11}\mathrm{W/cm}^2$. The injection UV pulse is centered at 291 nm ($11\omega_L$), with a duration of 3 cycles at $-0.25T_L$ delay, shown in the inset of Fig. \ref{HHG}(b). Fig. \ref{HHG}(a) illustrates the harmonic spectra at various injection intensities. It can be seen that the cutoff in the IR scheme is $\omega_{\mathrm{cutoff}}^{\mathrm{IR}}\approx25\omega_L$, while that in the IR+UV scheme is extended to $\omega_{\mathrm{cutoff}}^{\mathrm{IR+UV}}\approx60\omega_L$ in the secondary plateau.

\begin{figure}[htb]
	\includegraphics[scale=0.75]{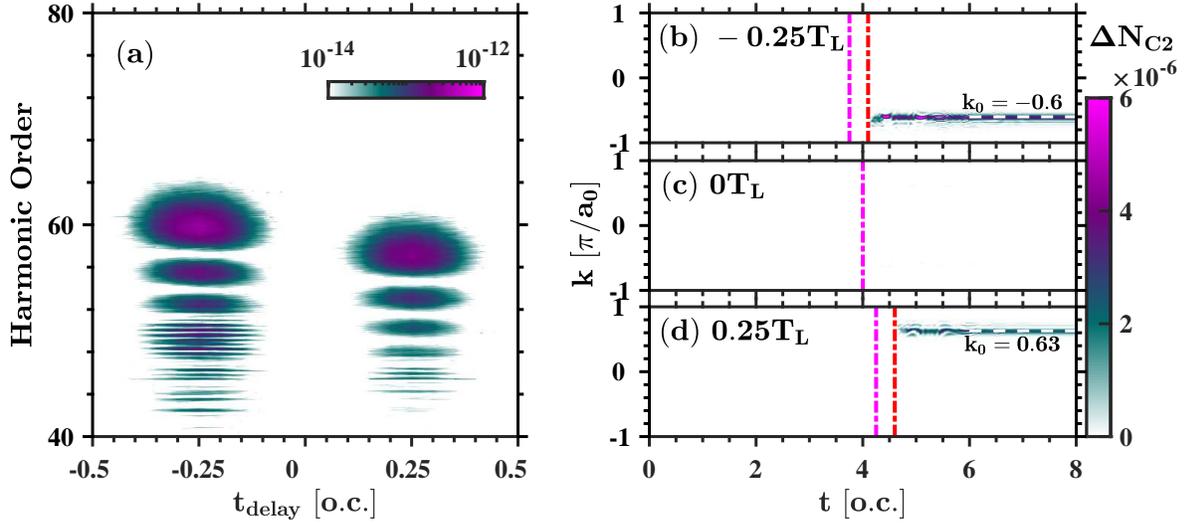}
	\centering
	\caption{ \label{Rhokt}
		(a) Dependence of harmonic spectrum on two-color delays, with the second plateau presented in logarithmic scale; Population difference of band C2 $\Delta N_{\mathrm{C2}}(t,k)$ between the two schemes at the delay of (b) $-0.25T_{\mathrm{L}}$, (c) $0T_{\mathrm{L}}$ and (d) $0.25T_{\mathrm{L}}$, with $\Delta N_{\mathrm{C2}}(t,k) = N_{\mathrm{C2}}^{\mathrm{IR+UV}}(t,k)-N_{\mathrm{C2}}^{\mathrm{IR}}(t,k)$.
		The generation of radiation in the extended plateau demands properly-delayed UV near the zero-crossing of electric field at $-0.25T_L$ or $0.25T_L$, and the highly-excited electrons on C2 locates far from the center at $k_0\approx-0.6\pi/a_0$ (white dashed line) and $0.63\pi/a_0$ (white dashed line), respectively, as predicted by the extended three-step model. The emergence of electrons on C2 (red dashed lines) is delayed to the injection instant (magenta dashed line) by $0.35T_L$ and proves the failure of channel 3 (see main text).
	}
\end{figure}

Though it is evident that emissions in the secondary region origin from significant population on band C2, there are actually three possible channels in the IR+UV setup:
\begin{enumerate}
    \item V1 to C1 and C1 to C2, both involve the absorption of one UV photon, hereinafter referred to as channel 1;
    \item V1 to C1 through absorption of one UV photon and C1 to C2 through tunnelling, hereinafter referred to as channel 2;
    \item V1 to C1 through tunnelling and C1 to C2 through absorption of one UV photon, hereinafter referred to as channel 3.
\end{enumerate}
Through detailed analysis of the intensity dependence and population variation, channel 2 is validated to contribute significantly to the population on C2. Firstly, as shown in Fig. \ref{HHG}(b), the linear dependence of the cutoff intensity on the injection intensity $I_{11\omega_L}$ implies that only one UV photon participates in the transition. In result, contribution from channel 1 can be easily omitted for the expected quadratic dependence on UV intensity. In channel 2, i.e. the extended three-step model, it can be predicted that injection at the extreme of electric field fails to populate electrons up to band C2, because additionally excited electrons still locate near the BZ center. Fig. \ref{Rhokt}(a) illustrates the dependence of secondary plateau on the injection delays. As can be seen, only injections near the delay of $-0.25T_L$ and $0.25T_L$ produce evident radiation. Moreover, Fig. \ref{Rhokt}(b)-(d) illustrate the evolution of population variation on band C2 at three typical delays, which show clearly that the excited electrons for the two delays locate initially at $k_0=-0.6\pi/a_0$ and $0.63\pi/a_0$, respectively, while the injection at zero delay fails to pump electrons to band C2. Finally, because the absorption of one UV photon is needed for transition from C1 to C2 in channel 3, obvious increasement of electrons on C2 should be seen at the instant of injection. However, as shown in Fig. \ref{Rhokt}(b)-(d), growth of population on C2 is delayed to the instant of UV injection by $0.35T_L$. As a result, the extended three-step model is proved to contribute mainly to the generation of harmonics in the secondary plateau.

\begin{figure}[htb]
\includegraphics[scale=0.75]{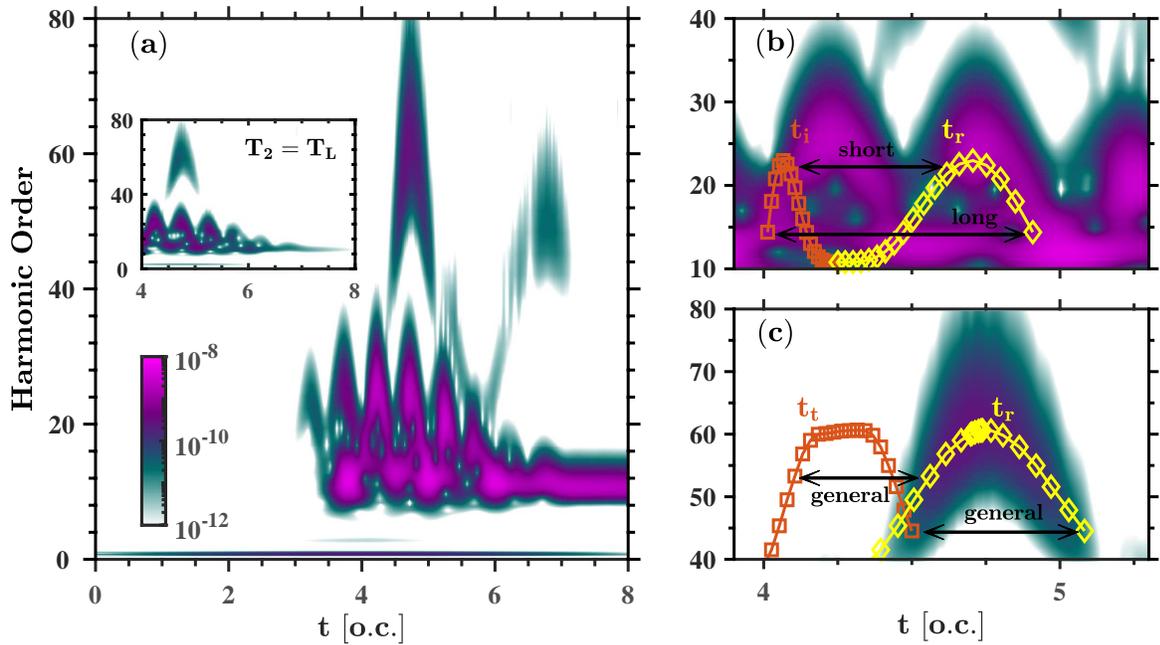}
\centering
\caption{ \label{WavJ}
(a) Time-frequency analysis of the two-color delay of $-0.25T_L$, with the result of shorter dephasing time $T_2=T_L$ presented in the inset. Strong dephasing leads to obvious suppress of of intrinsic emissions at both the minimal gap and $50\omega_L$ (gap between V1 and C2 at the crystal momentum of highly-excited electrons); Interband dynamics in the (b) first and (c) second plateaus employing acceleration theorem, with $t_i$ the ionization instant, $t_r$ the recombination time and $t_t$ the tunneling instant. Positive chirp in the primary plateau is linked to electrons that ionize later but recombine earlier, i.e. the so-called short trajectory, while that in the secondary plateau is related to electrons that tunnel and recombine earlier, i.e. the 'general' trajectory. 
}
\end{figure}

Though the transition channel has been validated, the multiband dynamics remain folded so far. Fig. \ref{WavJ}(a) shows the time-frequency analysis, where harmonic emission in the primary plateau exhibits clearly a period of half cycle of the fundamental field. The lack of clearness in the spectrum is mainly caused by long dephasing time, which also results in the quantum beat signal at the minimal gap after the end of laser field due to strong coherence of Bloch states near the gap\cite{Korbman2013quantum}. With the decrease of dephasing time down to $T_L$, the beat signal gradually disappears as shown in the inset. Meanwhile, significant emission at $t=4.75T_L$ can also be seen clearly in the second plateau and is delayed by approximately one optical cycle to the injection instant at $3.75T_L$. Surprisingly, slightly weakened harmonics are also generated at $t=6.75T_L$, which is emitted by the coherent Bloch states at the crystal momentum ($-0.6\pi/a_0$) of excited electrons, see Fig. \ref{Rhokt}(b). The intrinsic emission can also be suppressed by shorter dephasing time as shown in the inset.

We then examine the subcycle electron dynamics in both the first and second plateau. Within the acceleration theorem, the spacial displacement of electron hole pairs can be determined by
\begin{equation} \label{acc_theory1}
R_{e,h}(t) = \int^{t_r}_{t_i}\nabla_{\mathbf{k}} E(\mathbf{k}_0+A(\tau)-A(t_i)) d\tau,
\end{equation}
where $\mathbf{k}_0$ is the initial crystal momentum of excited electrons, $t_i$ and $t_r$ is the ionization and recombination instant, respectively. $t_r$ and energy of harmonic $\omega$ is given by
\begin{eqnarray}
R_e - R_h =& 0,\\ \label{acc_theory2}
E_e\left[\mathbf{k}_0+A(t_r)-A(t_i)\right] - E_h\left[\mathbf{k}_0+A(t_r)-A(t_i)\right] =& \hbar \omega.\label{acc_theory3}
\end{eqnarray}
Taking ionization near $4T_L$ as an example, Fig. \ref{WavJ}(b) shows the dependence of harmonic energy and recombination time on ionization time by solving Eq. (\ref{acc_theory1})-(\ref{acc_theory3}) with $\mathbf{k}_0=0$. It can be seen that electrons ionized right after the field extreme ($4T_L\le t\le4.06T_L$) recombine at delays larger than $\sim 0.7T_L$, while those ionized later ($4.06T_L< t\le4.25T_L$) recombine earlier, i.e. the so-called long and short trajectory respectively. Meanwhile, emissions from long trajectory exhibit negative chirp while those from short trajectory generate positive chirp. Similarly, the displacement of electrons in the extended three-step model is given by a convenient variation of Eq. (\ref{acc_theory1})
\begin{eqnarray} \label{acc_theory1_3}
R_{e}(t) = &\int^{t_t}_{t_i}\nabla_{\mathbf{k}} E_{\mathrm{C1}}(\mathbf{k}_0+A(\tau)-A(t_i)) d\tau + \nonumber \\
&\int^{t_r}_{t_t}\nabla_{\mathbf{k}} E_{\mathrm{C2}}(\mathbf{k}_0+A(\tau)-A(t_i)) d\tau,
\end{eqnarray}
with $t_t$ the tunneling instant from C1 to C2. Setting the initial crystal momentum to be $-0.6\pi/a_0$ as shown in Fig. \ref{Rhokt}(b) and the ionization time $t_i$ at the injection instant $3.75T_L$, Fig. \ref{WavJ}(c) shows the dependence of harmonic energy and recombination time on the tunneling instant. It can be seen that the recombination time and energy of harmonic emission shows nearly perfect match with the time-frequency analysis. Surprisingly, though emissions in the secondary plateau exhibit both positive and negative chirps, the chirps are not related to the so-called long or short trajectories in the interband dynamics of the first plateau. Actually, the positively and negatively chirped emission in the extended region is radiated by electrons from the `general' trajectory, where the electrons that ionize earlier recombine earlier with a delay of approximately $0.37\sim0.6T_L$. Varying the band dispersion gives a delay between $0.1T_L$ and $0.7T_L$, approaching that of the short trajectory. As a matter of fact, time interval between tunneling and recombination depends on the actual band dispersion of band C2, because it determines the group velocity of recombining electrons, indicated by Eq. (\ref{acc_theory1_3}). The sensitivity to band dispersion, on the other hand, can be possibly applied to reconstruct the band structure of higher-lying band.

\section{Conclusions}

In summary, we investigate the multiband dynamics of cut-off extension in the IR+UV scheme employing 1D SBE. Through linear scaling of cutoff intensity on injection intensity, delay sensitivity and time-dependent population difference, we solidify the foundation of extended three-step model in describing the origin of population increasement on high-lying conduction band. Surprisingly, deeper exploration into the time-frequency characteristics shows that though both positive and negative chirps coexists during the subcyle dynamics in the first and second plateau, the positive chirp in the first plateau is radiated by electrons that ionize later but recombine earlier while that in the second plateau is linked to electrons that tunnel earlier and recombine earlier. Our research may have implications for the generation of table-top attosecond sources and hold the promise of all-optical reconstruction of band structure beyond the lowest band edge. 

\ack

This study is supported by China National Key R\&D Program (Grant No. 2019YFA0307703), the NSF of China (Grant No. 12234020, 12274461 and 11904400), and the Major Research plan of NSF of China (Grant No. 91850201).

\section*{References}

\providecommand{\newblock}{}

\end{document}